\documentstyle[12pt]{article}
\begin{document}
\begin{flushright}
hep-th/0309245\\
SNB/September/2003
\end{flushright}
\vskip 3.7cm
\begin{center}
{\bf \Large {Notoph Gauge Theory as Hodge Theory}}

\vskip 3.5cm

{\bf R.P.Malik}
\footnote{ E-mail address: malik@boson.bose.res.in  }
\footnote{ Invited talk delivered in the {\bf International Workshop} on 
``Supersymmetries and Quantum Symmetries ''  
({\bf SQS '03}: 24-29 July, 2003) held
at BLTP, JINR, Dubna, Moscow Region, Russia in memory of {\bf Professor 
V. I. Ogievetsky} on the occasion of his 75th birth anniversary.}\\
{\it S. N. Bose National Centre for Basic Sciences,} \\
{\it Block-JD, Sector-III, Salt Lake, Calcutta-700 098, India} \\

\vskip 3.5cm

\end{center}

\noindent
{\bf Abstract:}
In the framework of an extended BRST formalism,  
it is shown that the four $(3 + 1)$-dimensional  (4D)
free Abelian 2-form (notoph) gauge theory presents
an example of a tractable field theoretical model for the Hodge theory.
\baselineskip=16pt

%\vskip 1cm

\newpage

\noindent
In their famous paper entitled ``Notoph and Its Possible Interactions''
(published in Russian [1] as well as in English [2]), Professor V. I. 
Ogievetsky (JINR, Dubna) and Professor I. V. Palubarinov (JINR, Dubna) coined
the name ``notoph'' (i.e. the opposite of the 1-form 
Abelian gauge field ``photon'')
for the antisymmetric ($B_{\mu\nu} = - B_{\nu\mu}$) 2-form
(i.e. $B = \frac{1}{2} (dx^\mu \wedge dx^\nu) B_{\mu\nu}$) Abelian
gauge field $B_{\mu\nu}$. This antisymmetric field has turned out to be quite
popular in the context of modern developments in the string theories and their
close cousins $D$-branes. For instance, this field appears very naturally in 
the supergravity multiplets [3], the excited states of quantized (super)string 
theory [4] and its existence is crucial for the anomaly cancellation in
the superstring theories [4,5]. Due to the presence of this field
in the  background, it turns out that the end points of the open strings
become {\it noncommutative} when they are trapped on the $D$-branes. Thus,
this field plays a crucial role in the modern upsurge of interest in
the noncommutative geometry [6]. Besides these connections, this field
and its corresponding gauge theory have been found to be relevant in
the realm of cosmic string theory, vortices in an incompressible 
and irrotational liquid, QCD, ``hairs'' on the black holes, etc. [7-10].
It generates an effective mass for the photon field in 4D through
a topological (i.e. celebrated $B \wedge F$) coupling 
term where the gauge invariance and mass co-exist
without any recourse to the Higgs mechanism [11,12]. In addition, 
as is well-known, this antisymmetric massless Abelian
field provides a dual description of the massless
scalar field theory in the four dimensions of spacetime.

The purpose of our present note is to add, yet another, novel feather
to the already richly feathered and beautifully
studded cap of 2-form Abelian gauge field theory as 
discussed in the above paragraph in the context of its various connections and
relevance to theoretical physics. We demonstrate that the free 4D Abelian
2-form gauge theory, in the framework of an extended 
Becchi-Rouet-Stora-Tyutin (BRST) formulation,
provides a beautiful field theoretical model for the Hodge theory where
all the de Rham cohomological operators $(d, \delta, \Delta)$, the
Hodge duality $*$ operation and the Hodge decomposition theorem
\footnote{ On a compact $D$-dimensional manifold without 
a boundary, any arbitrary
differential form $f_n$ of degree $n$
(with $ n = 0, 1, 2...; 0 \leq n \leq D$), can be uniquely expressed as
$f_{n} = h_{n} + d e_{n-1} + \delta c_{n+1}$ where $(\delta)d$
($\delta = \pm * d *, d = dx^\mu \partial_\mu$ with $\delta^2 = d^2 = 0$)
are the (co-)exterior derivatives which define the Laplacian
operator $\Delta = (\delta + d)^2 = \{\delta, d\}$. Here $*$ is the
Hodge duality operator on the manifold and $h_n$ is the harmonic
form (i.e. $ \Delta h_n = d h_{n} = \delta h_n = 0$).
It will be noted that $d$ and $\delta$
raise and lower the degree of a form by one, respectively. These cohomological
operators satisfy the algebra: $ d^2 = \delta^2 = 0, \Delta = \{d , \delta \},
[\Delta , d ] = [\Delta, \delta ] = 0$. This shows that $\Delta$ is
the Casimir operator for the whole algebra [13-15].} etc., can be
expressed in terms of the local, covariant and continuous symmetry 
transformations (and the corresponding generators) for the
BRST- and dual(co)-BRST invariant Lagrangian density of the theory.
In the realm of BRST cohomology, it is a common folklore to identify
the conserved ($\dot Q_{b} = 0$) and nilpotent $(Q_b^2 = 0)$
BRST charge $Q_b$ (that generates a local, covariant, continuous
and nilpotent symmetry transformations for the Lagrangian density
of a given gauge theory) with the nilpotent
($d^2 = 0$) exterior derivative $d =
dx^\mu \partial_\mu $ of differential geometry. It has been a
{\it long-standing problem} to explore the possibility of identifying
$\delta, *, \Delta$ with some interesting symmetry properties of the
Lagrangian density of a given gauge theory so that one can obtain
a field theoretical model for the Hodge theory. To demonstrate
that the notoph gauge theory does provide such an interesting 4D
model for the Hodge theory, we begin with the following (anti-)BRST
invariant Lagrangian density $({\cal L}_{b})$ [16,17]
$$
\begin{array}{lcl}
{\cal L}_{b} &=& \frac{1}{12} H_{\mu\nu\kappa} H^{\mu\nu\kappa}
+ B^\mu (\partial^\nu B_{\mu\nu} - \partial_\mu \phi_1)
- \frac{1}{2} B^\mu B_\mu
- \partial_\mu \bar \beta \partial^\mu \beta \nonumber\\
&+&
(\partial_\mu \bar C_\nu - \partial_\nu \bar C_\mu)\; (\partial^\mu C^\nu)
+ \rho\; (\partial \cdot C + \lambda) 
+ (\partial \cdot \bar C + \rho)\; \lambda,
\end{array}\eqno(1)
$$
where $B_\mu, \phi_1$ are the bosonic auxiliary fields, $\rho, \lambda$
are the scalar fermionic ($\rho^2 = \lambda^2 = 0, \rho \lambda 
+ \lambda \rho = 0$) auxiliary (ghost) fields, $(\bar C_\mu)C_\mu$ are
the fermionic $(C_\mu^2 = \bar C^2_\mu = 0, C_\mu \bar C_\nu + \bar C_\nu
C_\mu = 0$ etc.) vector (anti-)ghost fields and $(\bar \beta)\beta$
are the bosonic (anti-)ghost fields. Here the kinetic energy term
$(\frac{1}{12} H_{\mu\nu\kappa} H^{\mu\nu\kappa})$ is constructed 
(i.e. $H_{\mu\nu\kappa} = \partial_\mu B_{\nu\kappa} +$ cyclic terms)
from the antisymmetric gauge field $B_{\mu\nu}$ of the 2-form field
$(B = \frac{1}{2} (dx^\mu\wedge dx^\nu) B_{\mu\nu})$ when it is acted upon by
the exterior derivative $d$ to generate a 3-form $H$ (i.e.$ H = d B
= \frac{1}{3!} (dx^\mu \wedge dx^\nu \wedge dx^\kappa) H_{\mu\nu\kappa}$).
The above quadratic kinetic energy term can be linearized by introducing
a massless ($\Box \phi_2 = 0$) pseudo-scalar field $\phi_2$ and an
axial-vector
auxiliary field ${\cal B}_\mu$ as
\footnote{ We follow here
the convention and notations such that the flat metric
$\eta_{\mu\nu} = $ diag $(+1, -1, -1, -1)$ and the totally antisymmetric
Levi-Civita tensor ($\varepsilon_{\mu\nu\kappa\zeta}$) is chosen to
satisfy $\varepsilon_{\mu\nu\kappa\zeta} \varepsilon^{\mu\nu\kappa\zeta}
= - 4!, \varepsilon_{\mu\nu\kappa\zeta} \varepsilon^{\mu\nu\kappa\sigma}
= - 3! \delta^\sigma_\zeta$ etc. and $\varepsilon_{0123} = - \varepsilon^{0123}
= + 1, \varepsilon_{0ijk} = \epsilon_{ijk}$. Here the Greek indices 
$\mu, \nu.... = 0, 1, 2, 3$ correspond
to the 4D spacetime directions on the manifold and the Latin indices
$i, j, k...= 1, 2, 3$ stand for the space directions. It should be
noted that $(\partial \cdot C) = \partial_\mu C^\mu = \partial_0
C_0 - \partial_i C_i$ etc.}
$$
\begin{array}{lcl}
{\cal L}_{B} &=& \frac{1}{2} {\cal B}^\mu {\cal B}_\mu - {\cal B}^\mu
(\frac{1}{2} \varepsilon_{\mu\nu\kappa\zeta} \partial^\nu B^{\kappa\zeta}
- \partial_\mu \phi_2)
+ B^\mu (\partial^\nu B_{\mu\nu} - \partial_\mu \phi_1)
- \frac{1}{2} B^\mu B_\mu \nonumber\\
&-& \partial_\mu \bar \beta \partial^\mu \beta +
(\partial_\mu \bar C_\nu - \partial_\nu \bar C_\mu)\; (\partial^\mu C^\nu)
+ \rho\; (\partial \cdot C + \lambda) 
+ (\partial \cdot \bar C + \rho) \;\lambda.
\end{array}\eqno(2)
$$
The obvious point, at this stage, is the fact 
that the massless ($\Box \phi_{2} =
\Box \phi_1 = 0$) (pseudo-)scalar fields $(\phi_2)\phi_1$ behave with 
each-other in the same manner as
the dual antisymmetric field $(\frac{1}{2} \varepsilon_{\mu\nu\kappa\zeta}
B^{\kappa\zeta})$ behaves with $(B_{\mu\nu})$ in 4D. Furthermore, it is clear
that for the Lagrangian density (2) to possess simultaneously
the (anti-)BRST and (anti-)co-BRST symmetries, {\it the
$B_{\mu\nu}$ field and its dual $\phi_2$ field are required to be present 
together in it}. Under the following (i) 
local, covariant, continuous and off-shell nilpotent 
$(s_{b}^2 = 0)$ BRST transformations
$$
\begin{array}{lcl}
&& s_{b} B_{\mu\nu} = (\partial_\mu C_\nu - \partial_\nu C_\mu), \qquad
s_{b} C_\mu = \partial_\mu \beta, \qquad s_{b} \beta = 0, \qquad
s_b \bar C_\mu = B_\mu, \nonumber\\
&& s_{b} \phi_1 = - \lambda, \qquad s_b \bar \beta = \rho, \qquad
s_b (H_{\mu\nu\kappa}, \phi_2, {\cal B}_\mu, \rho, B_\mu, \lambda) = 0,
\end{array}\eqno (3)
$$
and (ii) local, covariant, continuous
off-shell nilpotent $(s_d^2 = 0$) dual(co-)BRST transformations
$$
\begin{array}{lcl}
&& s_{d} B_{\mu\nu} = \varepsilon_{\mu\nu\kappa\zeta} 
\partial^\kappa \bar C^\zeta, \qquad
s_{d} \bar C_\mu = - \partial_\mu \bar \beta, \qquad s_{d} \bar \beta = 0, 
\qquad
s_d  C_\mu = {\cal B}_\mu, \nonumber\\
&& s_{d} \phi_2 = - \rho, \qquad s_d  \beta = \lambda, \qquad
s_d (\partial^\mu B_{\mu\nu}, \phi_1, {\cal B}_\mu, \rho, B_\mu, \lambda) = 0,
\end{array}\eqno (4)
$$
the above Lagrangian density (2) transforms to some
 total derivatives. Furthermore,
it can be checked that, under the following discrete symmetry transformations
$$
\begin{array}{lcl}
&& \phi_{1} \rightarrow \pm i \phi_2, \qquad \phi_2 \rightarrow \mp i \phi_1,
\qquad B_\mu \rightarrow \pm i {\cal B}_\mu, \nonumber\\
&& {\cal B}_\mu \rightarrow \mp i B_\mu, \qquad
B_{\mu\nu} \rightarrow \mp \frac{i}{2} \varepsilon_{\mu\nu\kappa\zeta}
B^{\kappa\zeta}, \nonumber\\
&& C_\mu \rightarrow \pm i \bar C_\mu, \qquad \beta \rightarrow \mp 
i \bar \beta, \qquad \bar\beta \rightarrow \pm i \beta, \nonumber\\
&& \rho \rightarrow \pm i \lambda, \qquad \lambda \rightarrow \pm i \rho,
\qquad \bar C_\mu \rightarrow \pm i C_\mu,
\end{array} \eqno(5)
$$
the Lagrangian density (2) remains invariant. This re-establishes the fact that
the pairs ($\phi_1, \phi_2$), ($B_\mu, {\cal B}_\mu)$ are 
{\it dual} to each-other
on the same footing as $(B_{\mu\nu})$ and $(\frac{1}{2} 
\varepsilon_{\mu\nu\kappa\zeta} B^{\kappa\zeta})$ are dual to each-other.
The key
and distinguished features to be noted, at this stage, are (i) the ghost
part of the Lagrangian density ${\cal L}_{g} =
- \;\partial_\mu \bar \beta \;\partial^\mu \beta +
(\partial_\mu \bar C_\nu - \partial_\nu \bar C_\mu)\; (\partial^\mu C^\nu)
+ \rho\; (\partial \cdot C + \lambda) + (\partial \cdot \bar C 
+ \rho)\; \lambda$,
remains invariant in {\it itself} due to the discrete symmetry transformations 
on the (anti-)ghost fields of (5). On the contrary, the kinetic energy
and the gauge-fixing terms ${\cal L}_{k.e.} + {\cal L}_{g.f.} =
 \frac{1}{2} {\cal B}^\mu {\cal B}_\mu - {\cal B}^\mu
(\frac{1}{2} \varepsilon_{\mu\nu\kappa\zeta} \partial^\nu B^{\kappa\zeta}
- \partial_\mu \phi_2) + B^\mu (\partial^\nu B_{\mu\nu} - \partial_\mu \phi_1)
- \frac{1}{2} B^\mu B_\mu$
exchange with each-other under the discrete transformations (5). (ii)
The substitution of the
discrete transformations on the {\it ghost fields} (cf. (5)) lead to the
derivation of the off-shell nilpotent $(s_{ab}^2 = s_{ad}^2 = 0$) anti-BRST
and anti-co-BRST symmetry transformations from (3) and (4). (iii) Under
the (anti-)BRST and (anti-)co-BRST transformations, it is the kinetic
energy term (more precisely $H_{\mu\nu\kappa}$ in $H = d B$) and
the gauge-fixing term (more accurately 
$\partial_\mu B^{\mu\nu}$ in $\delta B = \partial_\mu B^{\mu\nu}
dx_\nu$) remain invariant, respectively. (iv) It can be seen that
anticommutator of (anti-)BRST and (anti-)co-BRST symmetry transformations
is not zero and it defines a bosonic symmetry $s_{w} = \{s_b, s_d \} =
\{s_{ad}, s_{ab} \}, s_w^2 \neq 0$) (see, e.g., [17] for details)
$$
\begin{array}{lcl}
&&s_{w} B_{\mu\nu} = \partial_\mu B_\nu - \partial_\nu B_\mu +
\varepsilon_{\mu\nu\kappa\zeta} \partial^\kappa B^\zeta, \qquad
s_{w} C_\mu = \partial_\mu \lambda, \nonumber\\
&& s_{w} (\phi_1, \phi_2, B_\mu, {\cal B}_\mu, \rho, \lambda, \beta,
\bar\beta) = 0, \qquad\;\; s_w \bar C_\mu = - \partial_\mu \rho,
\end{array} \eqno(6)
$$
under which the Lagrangian density (2) transforms to a total derivative.
The above bosonic symmetry transformation $s_w$ is characterized by the fact
that either the (anti-)ghost fields do not transform or they transform
up to a vector gauge transformation. (v) There exist a global scale symmetry
in the theory that corresponds to the ghost symmetry transformation. The
infinitesimal version of this symmetry transformation ($s_g$) is
$$
\begin{array}{lcl}
&& s_g B_{\mu\nu}
 = s_{g} B_\mu = s_g {\cal B}_\mu = s_{g} \phi_1 = s_g \phi_2 = 0,
\qquad s_g C_\mu = \Sigma C_\mu, \qquad s_g \bar C_\mu = - \Sigma \bar C_\mu,
\nonumber\\
&& s_g \rho = - \Sigma \rho, \qquad s_g \lambda = \Sigma \lambda, \qquad
s_g \beta = 2 \Sigma \beta, \qquad s_g \bar \beta = - 2 \Sigma \bar \beta,
\end{array}\eqno(7)
$$
where $\Sigma$ is a global (spacetime independent)
parameter. There are no ghost scale transformations on the bosonic
fields $B_{\mu\nu}, B_\mu. {\cal B}_\mu, \phi_1, \phi_2$ because they carry
ghost number equal to zero. The $\pm 1$ signs for the pairs 
($C_\mu, \bar C_\mu$)
and ($\lambda, \rho)$ are due to their ghost numbers being $\pm 1$. On
the contrary, the factor of $\pm 2$ in the above transformations
for the pair $(\beta, \bar \beta)$ is
due to their ghost number being $\pm 2$. Thus, there are six local,
covariant and continuous symmetries in the theory out of which
four are nilpotent (i.e. $s_{9)b}^2 = s_{(a)d}^2 = 0$).

Now we would like to dwell a bit on the discrete symmetry transformations
(5). These transformations correspond to the Hodge duality $*$ operation of
the differential geometry in the following way. It can be checked the
{\it an interplay} between the nilpotent continuous symmetries 
($s_{(a)b}, s_{(a)d}$) and the discrete
symmetry (5) leads to the following relationship
$$
\begin{array}{lcl}
s_{(a)d} \; \Phi = \pm \; * \; s_{(a)b}\; * \; \Phi,
\end{array}\eqno(8)
$$
where $\Phi$ stands for the generic fields of the Lagrangian density (2)
and $*$ corresponds to the transformations in (5).
The $\pm$ signs in the above are dictated by the general requirement for a
duality invariant theory (see, e.g., [18] for details). Reduced to
its simplest form, this requirement is equivalent to the implementation 
of a couple of successive $*$ (cf. equation (5))
operation on a specific field. The resulting
signature in this operation leads to the determination of signs in (8). In
mathematical terms, this requirement, for the Lagrangian density (2), is
as follows
$$
\begin{array}{lcl} 
*\; ( \; *\; \Phi) = \pm\; \Phi \;\;\;\rightarrow\;\;\;
*\; ( \; *\; B) = = + \; B, \qquad
*\; ( \; *\; F) = -\; F, 
\end{array} \eqno(9)
$$
where $B = \beta, \bar \beta, \phi_1, \phi_2, B_{\mu\nu}, B_\mu. {\cal B}_\mu$
and $F = \rho, \lambda, C_\mu, \bar C_\mu$
and $*$ corresponds to the discrete transformations in (5). 
At this juncture, we compare and contrast between relationship (8) and the 
similar kind of relationship that exists in 
differential geometry. These are (i) the (co-)exterior derivatives
$(\delta)d$ defined on a given $D$-dimensional manifold are related
(i.e. $\delta = \pm * d *$) in exactly the same manner as $s_{(a)d}$ and
$s_{(a)b}$ are related with each-other in (8). (ii) For the notoph
gauge theory, the bosonic- and fermionic fields correspond to $\pm$
signs in (8), respectively (as is evident from our discussions before
equation (9)). The $\pm$ signs in $\delta = \pm * d *$
is dictated by the dimensionality of the manifold. For instance, for
an even dimensional manifold, there is always a negative sign on the
r.h.s. of the relationship: $\delta = - * d *$ (see, e.g., [13] for
details).

As noted earlier, there are six continuous symmetries for the 4D notoph
gauge theory described by the (anti-)BRST- and (anti-)co-BRST invariant
Lagrangian density (2). According to Noether's theorem, these
symmetries lead to the derivation of a set of six local conserved charges
(viz. $Q_{(a)b}, Q_{(a)d}, Q_{w}, Q_{g}$). The four conserved
and nilpotent ($Q_{(a)b}^2 = Q_{(a)d}^2 = 0$)
charges $Q_{(a)b}$ and $Q_{(a)d}$ (corresponding to the cohomological
operators $d$ and $\delta$) generate the 
(anti-)BRST and (anti-)co-BRST symmetries. A bosonic symmetry generator
$Q_{w} = \{Q_{b}, Q_{d} \} = \{ Q_{ad}, Q_{ab} \}$ (that corresponds
to the Laplacian operator) generate a bosonic symmetry transformation $s_w$.
Finally, the conserved charge $Q_g$ generates the scale transformations
for the (anti-)ghost fields (cf. eqn. (7)). These generators, analogous
to the algebra obeyed by the de Rham cohomological operators, follow the
algebra as is given below
$$
\begin{array}{lcl} 
&& Q_{(a)b}^2 = Q_{(a)d}^2 = 0, \qquad Q_{w} = \{ Q_{d}, Q_b \} =
\{Q_{ab}, Q_{ad} \}, \qquad \{Q_{d}, Q_{ad} \} = 0, \nonumber\\
&& i [ Q_g, Q_{b(ad)} ] = + Q_{b(ad)}, \qquad i [Q_g, Q_{d(ab)} ] - Q_{d(ab)},
\qquad \{ Q_{b}, Q_{ad} \} = 0,\nonumber\\
&& [ Q_w, Q_r ] = 0, \quad r = b, ab, d, ad, w, g, \qquad
\{ Q_b, Q_{ab} \} = 0, \quad \{ Q_d, Q_{ab} \} = 0,
\end{array} \eqno(10)
$$
which shows that $Q_w$ is the Casimir operator for the whole algebra. The
above algebra implies that if the ghost number 
($ i Q_g |\Psi>_n = n |\Psi>_n$) of a state $|\Psi>_n$ is $n$, then the
following relations ensue
$$
\begin{array}{lcl} 
&&i Q_g Q_{b(ad)}\; |\Psi>_n = (n + 1) \; |\Psi>_n, \nonumber\\
&&i Q_g Q_{d(ab)}\; |\Psi>_n = (n - 1) \; |\Psi>_n, \nonumber\\
&&i Q_g Q_{w}\; |\Psi>_n =\; (\;n\;) \; |\Psi>_n,
\end{array} \eqno(11)
$$
which imply that the states $Q_{b(ad)} |\Psi>_n, Q_{d(ab)} |\Psi>_n$
and $Q_{w} |\Psi>_n$ have the ghost number equal to $(n + 1), (n - 1)$
and $n$ respectively.
The above equation also implies that the charges $Q_{b(ad)}$ and $Q_{d(ab)}$
raise and lower the
ghost number of state by {\it one} respectively. This property is analogous to
the operation of $d$ and $\delta$ on a differential form of degree $n$. Thus, 
we have a ``two-to-one'' mapping between the
conserved charges and the cohomological operators as:
$ Q_{b(ad)} \rightarrow d, Q_{d(ab)} \rightarrow \delta$ and
$Q_{w} = \{Q_b, Q_d \} = \{ Q_{ad}, Q_{ab} \} \rightarrow \Delta$.
The above mapping is primarily due to (i) the identification of the degree
of a form with the ghost number of a state, and
(ii) the algebra obeyed by the conserved charges in (10). 
With the above identifications, the analogue of the
celebrated Hodge decomposition theorem can be written, in the
quantum Hilbert space of states, as
$$
\begin{array}{lcl} 
|\Psi>_n = |\omega>_n + Q_{b(ad)} \; |\theta>_{n - 1} 
+ Q_{d(ab)}\; |\chi>_{n + 1},
\end{array} \eqno(12)
$$
where $n$ is the ghost number of an arbitrary state $|\Psi>_n$ and
$|\omega>_n$ is the harmonic state (i.e. $Q_{w} |\omega>_n = Q_{b(ad)}
|\omega>_n = Q_{d(ab)} |\omega>_n = 0$). Thus, we have established that
the notoph gauge theory, in the framework of an extended BRST formalism,
does provide
a beautiful field theoretical model for the Hodge theory where all
the de Rham cohomological operators correspond to certain specific symmetry
properties of the Lagrangian density (2).

To summarize, in our presentation, we have shown that the notoph gauge theory,
proposed by Ogievestsky and Palubarinov [1,2], provides an interesting field
theoretical model for the solution of the long-standing problem in the 
realm of BRST cohomology as, for this model, the analogues of the 
cohomological quantities $\delta, *$ and $\Delta$ do exist [17] in the language
of local, covariant and continuous symmetry properties of the Lagrangian
density (cf. equation (2))
 of this theory. Furthermore, it is shown [19] that this
field theoretical gauge theory is a model for the {\it quasi-topological}
field theory in 4D where the topological invariants do exist and they obey the
recursion relations that are characteristic features of an {\it exact}
topological field theory. However, the Lagrangian density and symmetric
energy-momentum of this theory are found {\it not} to be able to be expressed
as the sum of BRST and co-BRST exact quantities. In addition, this model
provides a fertile ground where Wigner's little group plays a very crucial
and decisive role in the discussion of the BRST cohomology and Hodge
decomposition theorem in the quantum Hilbert space of states [20]. 
There are many interesting directions that can be pursued later.
For instance, exploring the existence of the analogues of $\delta, *, \Delta$
for the {\it non-Abelian} version of the notoph gauge theory, 
study of its topological nature, its discussion in the framework of
superfield formulation, etc., are open problems connected with the
generalizations of the above 2-form Abelian gauge theory. Thus,
we conclude that the notoph gauge theory 
and its possible generalizations are expected to be endowed with
a very rich innate mathematical and physical structures which will enable
them to remain the hot-bed of theoretical research in years to come. In
particular, the higher dimensional theories will earnestly require them.

I wish to wrap up this presentation with a personal note of tribute to
Professor V. I. Ogievetsky. I had the privilege and honour to work in his
group ``Problems in Supersymmetry'' at JINR, Dubna during the later
part of his life. His mere presence at JINR used to be an awe-inspiring
experience for people of my generation. I owe a great deal to him for
my {\it intellectual-lineage} as I worked with Prof. E. Ivanov, 
Prof. S. Krivonos and Prof. A. Isaev at BLTP, JINR, Dubna. Prof. Ivanov
is a direct student of Prof. Ogievetsky and Prof. Krivonos and Prof.
Isaev have a whole range of research papers with Prof. Ivanov. Thus,
I would remain grateful to Prof. Ogievetsky and others (who are his
student and grand-students), 
throughout my life, for my intellectual debt.

\end{document}